\documentclass[aps,prl,a4paper,twocolumn,superscriptaddress,english,longbibliography]{revtex4-1}
\usepackage{amsmath,amssymb}
\usepackage{graphicx,epsfig,sidecap,wasysym}
\usepackage{color,epstopdf}
\usepackage{mathtools}
\usepackage[colorlinks,bookmarks=false,citecolor=blue,linkcolor=blue,urlcolor=blue]{hyperref}
\usepackage{float}

\def\be{\begin{equation}}
\def\ee{\end{equation}}

\def\bea{\begin{eqnarray}}
\def\eea{\end{eqnarray}}

\begin{document}
 \title{ Many-body localization of 1D disordered impenetrable two-component fermions}
	                                                                                                                                                                                                                                                                                                                                                                                                                                                                                                                                                                                                                                                                                                                                                                                                                                                                                                                                                                                                                                                                                                                                                                                                                                                                                                                                                                                                                                                                                                                                                                                                                                                                                                                                                                                                                                                                                                                                                                                                                                                                      
\author{M.\,S. Bahovadinov}
\email{mbakhovadinov@hse.ru}
\affiliation{Russian Quantum Center, Skolkovo, Moscow 143025, Russia}
\affiliation{  Physics Department, National Research University Higher School of Economics, Moscow, 101000, Russia}

\author{D.\,V. Kurlov}
\affiliation{Russian Quantum Center, Skolkovo, Moscow 143025, Russia}

\author{B.\,L. Altshuler}
\affiliation{Physics Department, Columbia University, 538 West 120th Street, New York, New York 10027, USA}
\affiliation{Russian Quantum Center, Skolkovo, Moscow 143025, Russia}

\author{G.\,V. Shlyapnikov}
 \email{georgy.shlyapnikov@universite-paris-saclay.fr}
\affiliation{Russian Quantum Center, Skolkovo, Moscow 143025, Russia}
\affiliation{Moscow Institute of Physics and Technology, Inst. Lane 9, Dolgoprudny, Moscow Region 141701, Russia}
\affiliation{Universit\'e Paris-Saclay, CNRS, LPTMS, 91405 Orsay, France}
\affiliation{Van der Waals-Zeeman Institute, Institute of Physics, University of Amsterdam,Science Park 904, 1098 XH Amsterdam, The Netherlands}

 \begin{abstract} 
  We study effects of disorder on eigenstates of 1D two-component fermions with  infinitely strong Hubbard repulsion. We demonstrate that the spin-independent (potential) disorder reduces the problem to the one-particle Anderson localization taking place at arbitrarily weak disorder. In contrast, a random magnetic field can cause reentrant many-body localization-delocalization transitions. Surprisingly weak magnetic field destroys one-particle localization caused by not too strong potential disorder, whereas at much stronger fields the states are many-body localized. We present numerical support of these conclusions.
 \end{abstract}

\maketitle
\noindent{{\it Introduction}.} 
Many-body localization (MBL) emerges from an interplay between disorder and interparticle interactions, extending the phenomenon of Anderson localization (AL) to many-body systems~\cite{Altshuler2006,Oganesyan_Huse_2007}. Numerous works revealed MBL-caused vanishing steady transport, absence of thermalization, area-law scaling of entanglement  entropy, etc.~\cite{Berkelbach2010,Barisic2016,YBLev2015,Herbrych2017,Pal_Huse_2010,YBLev2014,Serbyn2015,Luitz2016,De_Luca_2013,De_Luca_2014,Nayak_2013,Serbyn2013,Gritsev2019,Serbyn2016,Papic2013,Rademaker2016,Chandran2015}. For reviews, see e.g.~\cite{LuitzYBLev,AbaninReview,AletReview}.  

   MBL has been extensively studied in systems one-dimensional (1D) spinless fermions and various spin chain models~\cite{Alet2015,Kudo2018,LuitzMBL2016,BurinXY}. These models possess only a single local degree of freedom (LDoF), which is not the case in more realistic systems. For instance, the Fermi-Hubbard model possesses two LDoF, charge and spin, both of which can be coupled to disorder. Recently, the disordered 1D Hubbard model in the regime of a finite interaction strength was studied numerically~\cite{Rigol2015,Delande2018,Mierze2017,Lemut2017, Mierze2018,Kozarzewski,Filippone}. The results suggest that a sufficiently strong random potential localizes the charge degree of freedom, whereas spin excitations apperently exhibit a subdiffusive transport, i.e. remain delocalized~\cite{Protopopov2019,Znidaric2016,Protopopov2017,PrelovsekPMBL}. 
 A symmetric situation is observed in a sufficiently strong random magnetic field coupled to spins: spin excitations are localized whereas charge excitations are extended~\cite{Rachel2019}.
 
This Letter is devoted to the 1D Hubbard model for two-component fermions in the limit of infinitely strong  on-site repulsion. Theoretical considerations supported by the numerical analysis allowed us to demonstrate that arbitrarily weak potential disorder localizes both charge and spin excitations. Indeed, each site of the 1D lattice in this limit can be in 3 (rather than 4 as in the case of finite on-site interaction) distinct states: empty and single occupied with 2 possible spin orientations. The absence of the double occupation reduces the possible dynamics to the motion of \textit{empty spaces} (holons). The holons behave themselves like non-interacting fermions~\cite{Abarenkova2000} even in the presence of a spin-independent disorder~\cite{DenisWork} and thus are subject to the conventional single-particle AL. A weak magnetic disorder causes an effective interaction between the holons, which can result in the delocalization, unless the disorder is sufficiently strong.  

{\it Model and its symmetries}. 
The Hamiltonian of the conventional 1D Hubbard model with an infinitely strong on-site repulsion on a ring with~$L$~sites is, 
\be \label{H_0}
	H_0 = - t \sum_{i=1}^{L} \sum_{\sigma = \uparrow, \downarrow} {\cal P} \left(c^{\dagger}_{i,\sigma}c_{i+1,\sigma}+ \text{H.c.} \right) {\cal P}.
\ee
Here  $c_{j,\sigma}$ is the annihilation operator of a fermion in the spin state $\sigma$ in the site $j$, and $t$ is the nearest neighbor hopping amplitude. The operator ${\cal P} =\prod_i \left( 1- n_{i,\uparrow} n_{i, \downarrow} \right)$  with $n_{i,\sigma} = c^{\dag}_{i,\sigma}c_{i,\sigma}$, projects out the states with doubly-occupied sites. Below we impose periodic boundary conditions ($c_{L+1}=c_{1}$) and set~$t=1$. The integrable Hamiltonian (\ref{H_0}) was fully analyzed by means of the Bethe Ansatz~\cite{Hubbard_book}. 

The disorder is represented by the additional term in the Hamiltonian:
\be \label{H_D}
	H_{D}= \sum_{i =1}^{L} \varepsilon_{i} \left( n_{i,\uparrow} + n_{i,\downarrow}\right) + \sum_{i =1}^{L}  h_{i}  \frac{ n_{i,\uparrow} - n_{i,\downarrow}}{2},
\ee
 with $\varepsilon_i$ and $h_i$ being the random potential and magnetic field, respectively. We assume that both are uniformly distributed: $\varepsilon_i \in [-W, W]$ and $h_i \in [-B, B]$. 

The disorder breaks translational and SU(2) symmetries, but a number of symmetries are preserved. First, $H$ conserves the number of particles with a given spin $N_{ \left\lbrace \uparrow, \downarrow \right\rbrace } = \sum_{i} n_{i,\left\lbrace \uparrow, \downarrow \right\rbrace}$. Moreover, the infinite repulsion does not allow particles with opposite spins to exchange their positions, i.e. the spin pattern is conserved up to cyclical transmutations.  


{\it Localization measures.} Quantum states of the model~(\ref{H_0},\ref{H_D}) with a fixed set of conserved quantities form a ${\cal N_H}$-dimensional Hilbert space (${\cal N_H}<3^L$). We will analyze many-body eigenstates in the basis of states with a given occupation ($0$-empty, $\uparrow$,$\downarrow$-occupied by a particle with a particular spin direction) on each lattice site (computational basis): $|s \rangle= |s_1 \rangle \otimes |s_2\rangle \otimes ... \otimes |s_L \rangle$ and $|s_i\rangle \in \left\lbrace |0 \rangle, | \uparrow  \rangle, |\downarrow \rangle \right\rbrace$. 

Several quantitative measures have been proposed to characterize localization properties of the wavefunctions 
$\psi_\alpha(s)=\langle s  |  \alpha  \rangle $. Analysis of the scaling of the participation entropies $S_q$ with ${\cal N_H}$ allows one to determine fractal dimensions $D_q$:
\be
S_q=\frac{1}{q-1}\ln\left(\sum_{s=1}^{\cal N_H} |\psi_\alpha(s)|^{2q}\right) \xrightarrow{{\cal N_H} \rightarrow \infty} D_q \ln\left({\cal N_H}\right) .
\label{eq:Dq}
\ee
The eigenstates $|\alpha \rangle$ localized (LO) on a finite set of $|s \rangle$ have $S_q$ independent of ${\cal N_H}$ and thus $D_q=0$ for any $q>0$. Extended ergodic (EE) states with $|\psi_\alpha(s)|^2 \sim {\cal N_H}^{-1} $  are characterized by $D_q=1$. The states with $0<D_q<1$ are non-ergodic (NEE) albeit extended.
 The eigenstates which are supposed to be localized  when effective disorder is strong, can become extended as the disorder is reduced. For the conventional single-particle AL in 3D space at a critical strength of the disorder, the system undergoes a transition from LO to EE behavior. The NEE states are believed to exist only at the critical point. As for the MBL, there is a strong evidence of the existence of the NEE phase intermediate between LO and EE, implying two consecutive transitions LO $\rightarrow$ NEE and NEE $ \rightarrow$ EE at two different strengths of the disorder. 
 
 The fractal dimensions $D_q$ may be not very useful in the numerical identification of the LO $\rightarrow$ NEE transition: it is difficult to distinguish $D_q=0$ from $0<D_q \ll 1$  for a finite size system. More sensitive to this transition is the Kullback-Leibler divergence $KL$~\cite{Kullback1951,KravtsovCit,Pino2019}:\be
KL=\sum_{s=1}^{\cal N_H} | \psi_{\alpha}(s) |^2 \ln \left( \frac{|\psi_{\alpha}(s)|^2}{|\psi_{\alpha+1}(s)|^2} \right),
\label{eq:KL}
\ee  
(the states $|\alpha \rangle$  are supposed to be ordered in energy). The NEE states close to this transition can be viewed as the result of the hybridization of LO states. The LO states are not correlated in space and the ratio $| \frac{\psi_{\alpha}(s)}{\psi_{\alpha+1}(s) }|$ is exponentially large if $|\psi_{\alpha}(s)|$ is not negligible, i.e. KL diverges with ${\cal N_H}$. After the hybridization, NEE states $|\alpha \rangle$ and $|\alpha+1 \rangle$ involve mostly the same LO states. As a result $|\psi_{\alpha}(s)|$ and $|\psi_{\alpha+1}(s)|$ are strongly correlated with $| \frac{\psi_{\alpha}(s)}{\psi_{\alpha+1}(s) }| \sim  O(1)$ and $KL<\infty$. An abrupt change of $KL$ is therefore an indication of the LO to NEE transition.  

{\it Exact diagonalization.} 
In our exact diagonalization of the Hamiltonian $H = H_0 + H_D$ in the computational basis, we choose $L$ to be multiple of 4, $L \in \{12, 16, 20\}$, and confine ourselves to the sector with  $N_\uparrow=N_\downarrow=\frac{L}{4}$. The size of this sector is ${L\choose L/2}\times{L/2 \choose L/4}$.  The conservation of the \textit{spin pattern} further reduces the dimension of the Hilbert space to ${\cal N_H}={\cal Z}{{L}\choose{L/2}} $, where ${\cal Z}$ is the number of  distinct spin sequences connected by the cyclical transmutations for a given spin pattern. Below we only consider the Ne\'el spin pattern (${\cal Z}=2$) with the  corresponding ${\cal N_H} \in \{1848, 25740, 369512 \}$. We employ the shift-invert exact diagonalization algorithm to obtain $m \in \lbrace  20,20,100 \rbrace$~states from the central part of the energy spectrum. We average $KL$, $D_1$ and $D_2$ over these $m$ states and then perform averaging over $N_d$ realizations of the disorder (from $N_d \sim  10^4$ for the smallest system size up to $N_d \sim  10^2$ for the largest one). For later convenience, we parametrize the strengths of the random potential and magnetic field as,
\be \label{eq:disorderRho}
B = \rho_D \cos(\theta),\qquad W = \rho_D \sin(\theta),
\ee
where $\rho_D > 0$ is the total disorder strength.   
We carry out the calculations for various values of $\rho_D$ and $\theta$, ranging from the random potential ($\theta = \pi/2$ and $B=0$) to the random magnetic field ($\theta = 0$ and $W=0$).
 \begin{figure}[t]
\includegraphics[ width= \columnwidth]{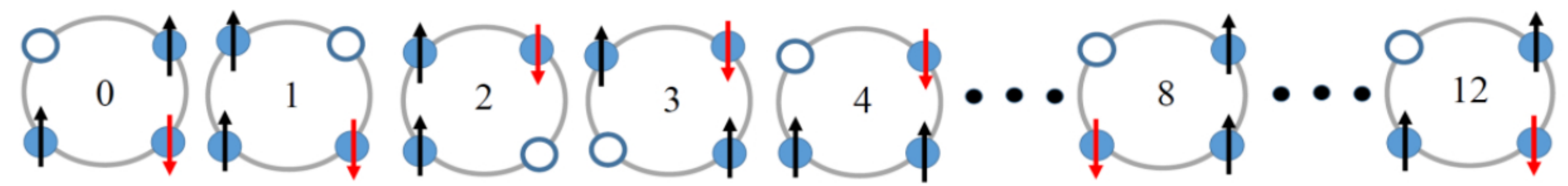}
\caption{ $L$=4 ring with one holon (empty site) and  spin pattern $|\uparrow \downarrow \uparrow \rangle $. A clockwise detour ($0 \circlearrowright 4 $) of the holon leads to a cyclical transmutation $|\uparrow \downarrow \uparrow \rangle \Rightarrow  |\downarrow \uparrow  \uparrow \rangle $. The system returns to the original state only after 3 detours: $|\uparrow \downarrow \uparrow \rangle \Rightarrow  |\downarrow  \uparrow \uparrow   \rangle  \Rightarrow  | \uparrow \uparrow \downarrow \rangle \Rightarrow | \uparrow \downarrow \uparrow \rangle $. Therefore, $\phi=\pm\frac{2\pi}{3}$. For a purely potential disorder the states 0, 4 and 8 have the same  potential energy, i.e. $\phi$ can be viewed as a quasimomentum of the system. } 
\label{fig:1}
\end{figure}


{\it Potential disorder.}
 In the Bethe Ansatz solution of the disorder-free problem governed by the Hamiltonian (\ref{H_0}) each holon is characterized by momentum $k_j$. The only difference of the holons from free spinless fermions on a ring is the quasi-periodic rather than periodic boundary condition that effects the quantization, $k_jL=2 \pi n_j + \phi$, $n_j \in \mathbb{Z}$.
Qualitatively, the extra phase $\phi $ can be understood as follows. The infinite Hubbard repulsion preserves the spin ordering but only up to cyclical transmutations. Each detour of a holon around the ring causes such a transmutation. To recover the original pattern the holon should make a number of detours equal to the number of non-equivalent cyclical transmutations of the pattern ${\cal Z}$ (see Fig.~\ref{fig:1}). For the spin pattern fixed up to the cyclical permutations the many-body states of the clean system $| \left\lbrace k_j \right\rbrace , \phi \rangle$ are characterized by the momenta of the holons $\left\lbrace k_j \right\rbrace$ and the total "quasimomentum" $\phi$ of the particles. The quantity $\phi$ should be a multiple of $2\pi/{\cal Z}$. For the Ne\'el pattern with an even number of particles we have ${\cal Z}=2$ allowing only $\phi=0$ and $\phi=\pi$.  
 
All of the above applies to the disordered model $H_0+H_D$  as long as the disorder is purely potential ($B=0,$  $ \theta=\frac{\pi}{2}$ in Eq.~(\ref{eq:disorderRho})). The potential disorder alters only one-particle states of the holons. The plane waves are substituted by the eigenfunctions of a particle subject to the random potential $\left\lbrace \varepsilon_i \right\rbrace$ with quasiperiodic boundary conditions. These one-particle states are known to be localized for arbitrarily small $W$. It is important to emphasize that as long as the random magnetic field is absent the disorder does not lead to any interaction between the holons and the occupation numbers of the one-particle states are conserved. The many-body wave function  is characterized by a macroscopically large set $\left\lbrace  \mu_j \right\rbrace_\phi$ instead of $\left\lbrace  k_j \right\rbrace$, where each $\left\lbrace  \mu_j \right\rbrace_\phi$ labels a localized one-particle wave function with a fixed $\phi$. This means that from the many-body point of view the problem remains integrable even in the presence of the potential disorder. The Bethe Ansatz based theory of the Hubbard model with potential disorder will be published elsewhere~\cite{DenisWork}.
\begin{figure}[t]
\includegraphics[ width= \columnwidth]{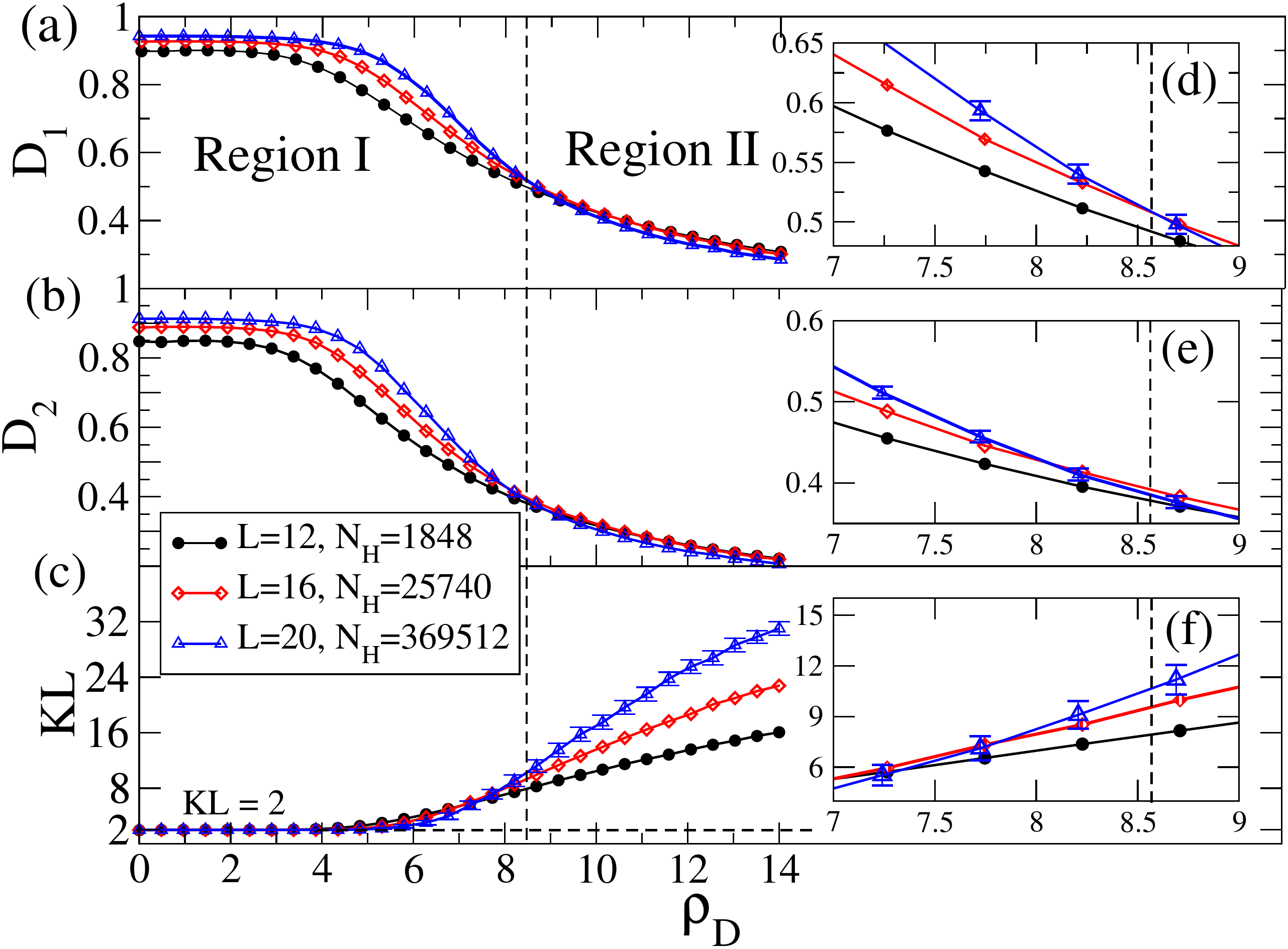}
\caption{  The fractal dimensions $D_1$ (a), $D_2$ (b), and the Kullback-Leibler divergence $KL$ (c) versus the total disorder strength~$\rho_D$ for the case of $\theta = 0$ (random magnetic field only, no potential disorder) for $L=12$ (black filled circles), $L=16$ (red open diamonds) and $L=20$ (blue filled triangles). Insets (d)-(f) show the data near the critical point $\rho_D\simeq 8.57$.  
 In Region I the states are extended, whereas in Region II they are localized. Error bars smaller than the symbol sizes are omitted. } 
\label{fig:2}
\end{figure}


{\it Random magnetic field.} In the presence of a spin-dependent disorder  ($\theta<\pi/2$ in Eq.~(\ref{eq:disorderRho})) a cyclical transmutation of the spin pattern caused by a holon detour changes the total energy (one can see, e.g., that the potential energies of the states 0, 4 and 8 in Fig.~\ref{fig:1} are all different). This violates the conservation of the 'quasimomentum'  $\phi$, i.e. gives rise to the off-diagonal matrix elements of the Hamiltonian, $\langle \left\lbrace \mu^\prime_j \right\rbrace, \phi^\prime | H |\left\lbrace \mu_j \right\rbrace, \phi \rangle \neq 0$.
We thus come to the conclusion that the random magnetic field plays a two-fold role. First, like the potential disorder, it tends to transform the plane waves into localized one-particle states. On the other hand, it hybridizes different  $|\mu_j \rangle$ - states and can cause the many-body delocalization in the $| \left\lbrace \mu_j \right\rbrace, \phi \rangle$ -basis.

Below we present the results of numerical studies of the MBL transition in the computational
basis $\left\lbrace | s \rangle \right\rbrace $ rather than in the $\left\lbrace  | \left\lbrace \mu_j \right\rbrace, \phi \rangle \right\rbrace$  basis, i.e. in the clean limit ($H=H_0$, $\rho_D=0$) the many-
body eigenstates $| \left\lbrace k_j \right\rbrace, \phi \rangle$  are apparently extended, whereas the one-particle localization of the
holon states $|\mu_j \rangle$ due to the potential disorder implies MBL in the computational basis. We
therefore have to check numerically the following predictions: i) If the disorder is purely magnetic, $\theta=0$ , the many-body states remain extended for a finite but weak enough disorder because the problem is not a 1D one-particle one. With increasing the randomness $\rho_D$ the system eventually becomes many-body localized, i.e. it undergoes the MBL transition; ii) For the purely potential disorder, $\theta=\pi/2$, the eigenstates
are always localized in the
$\left\lbrace |s\rangle \right\rbrace$ - basis. Sufficiently strong magnetic disorder can destroy this localization if $\rho_D$ is small enough.

Fig.~\ref{fig:2} presents the results of numerical diagonalization of the Hamiltonian (\ref{H_0}), (\ref{H_D})
in the absence of the potential disorder ($W=0$, $\theta=0$, $\rho_D=B$) for different values of the system size $L$.
One can see that for a weak disorder the fractal dimensions $D_{ 1,2}$ are close to the
ergodic value $1$. Moreover, they increase with $L$, i.e. it is plausible that the quantum states remain
not only extended but also ergodic. The deviations from $D_{1,2} =1$  can be viewed as a finite size
effect. If the disorder is strong, $\rho_D\gtrsim 8.6$, the fractal dimensions $D_{1,2}$ decrease as $L$ increases and it
is likely that $D_{1,2} \rightarrow 0$ for $L\rightarrow \infty$, which corresponds to the LO phase. At the transition between extended and LO regimes the curves  $D_{1,2}(\rho_D)$ for different $L$ should cross each other at a critical disorder $\rho_D=\rho^C_D$ and we find $ \rho^C_D \approx 8.57 $.

For intermediate magnetic disorder, $D_{1,2}$ increase with $L$ but are much smaller than 1 and decrease
as the disorder gets stronger. One may assume that in the limit $L\rightarrow \infty$ the fractal dimensions
remain between $1$ and $0$. This would signal the existence of the NEE phase.
However, it is impossible to exclude that $ D_{1,2} \xrightarrow{ L \rightarrow \infty}0$.
 The divergence $KL$ is sensitive to the transition to the LO state but not to the EE $\rightarrow$ NEE transition. We found that $KL$ is practically $L$-independent for $\rho_D<\rho^C_D$  as it should be if the states are extended. The $KL(L)$ - dependence at $\rho_D>\rho^C_D$ is another evidence of the LO phase at large
disorder. Note that with the accuracy limited by the finite size of the system the analysis of the
data for both the fractal dimensions $D_{1,2}$ and $KL$ yields the same value of the critical disorder.
\begin{figure}[t]
\includegraphics[ width= \columnwidth]{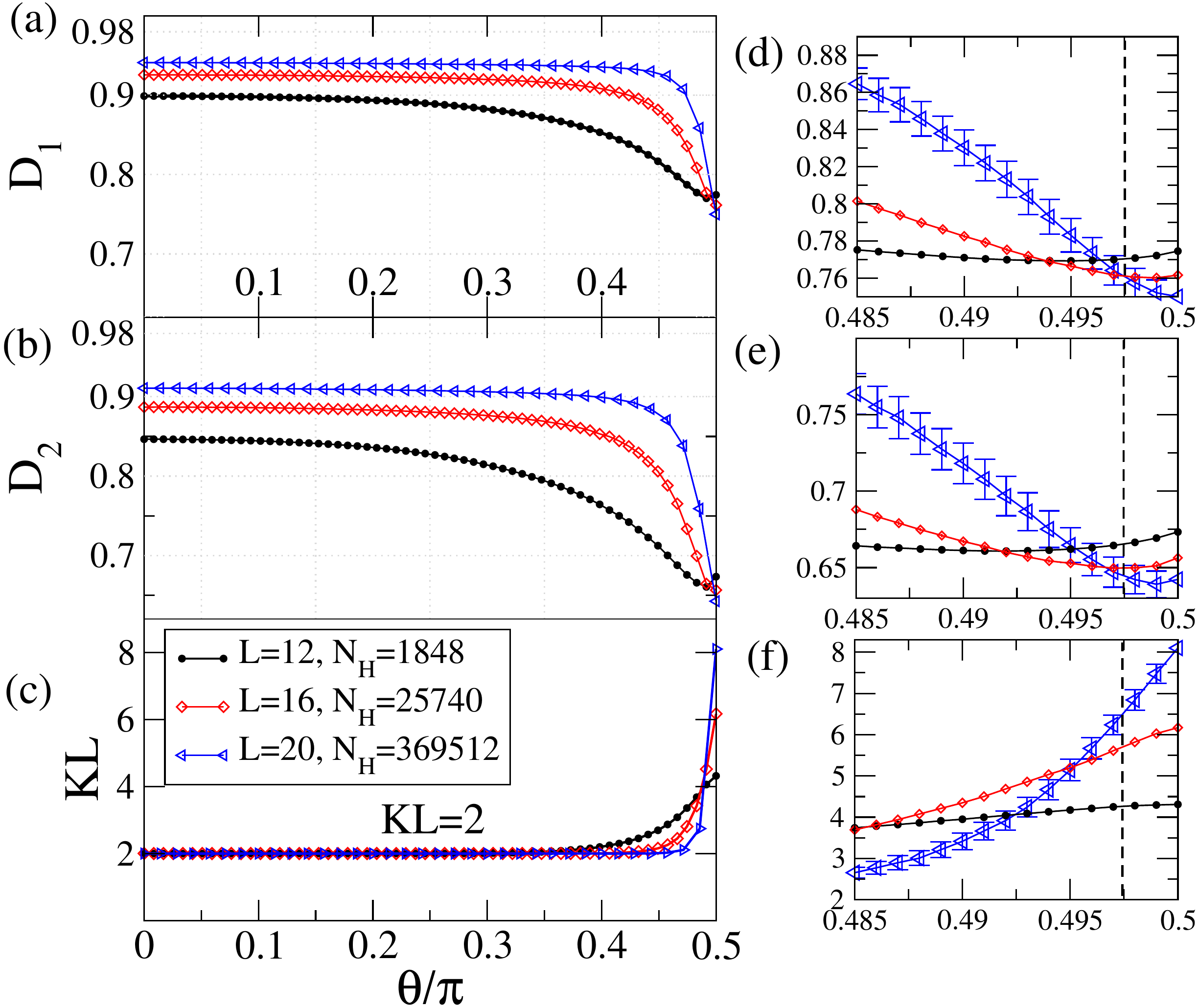}
\caption{  The fractal dimensions $D_{1}$ (a), $D_2$ (b), and the Kullback-Leibler divergence $KL$ (c) versus the angle~$\theta$ [see Eq. (\ref{eq:disorderRho})] at the total disorder strength $\rho_D = 2$ for $L=12$ (black filled circles), $L=16$ (red open squares), and $L=20$ (blue open triangles). Right panel (d)-(f) shows the data near $\theta=\pi/2$. Error bars smaller than the marker size are not shown.}  
\label{fig:3}
\end{figure}
\begin{figure}[t]
\includegraphics[width= \columnwidth ]{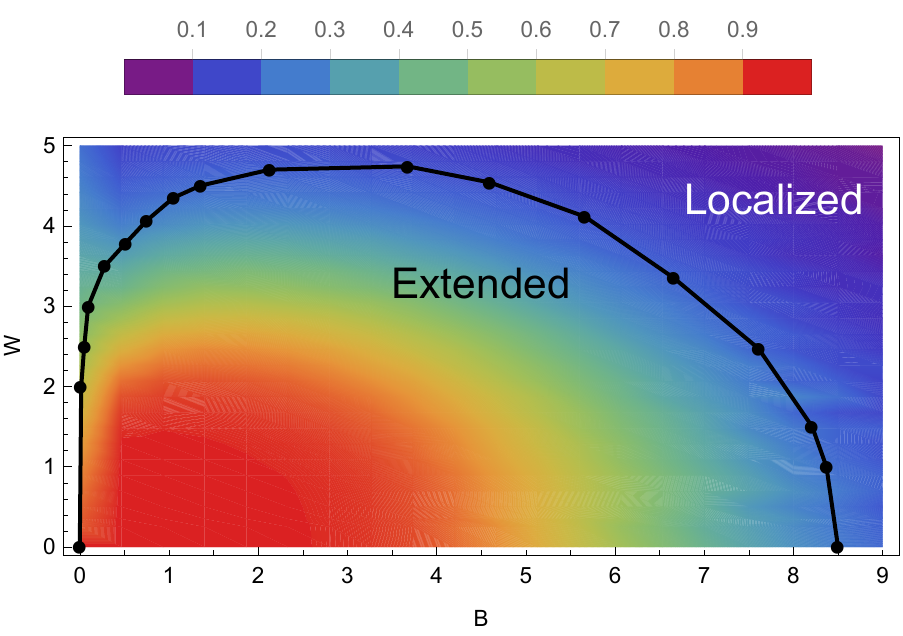}
\caption{ A sketch of the phase diagram in the $B-W$ space, based on the finite-size calculations of the fractal dimensions $D_1$. The colorbar indicates the value of $D_1$ for $L=16$. } 
\label{fig:4}
\end{figure}
 

{\it Two-component disorder.} Fig.~\ref{fig:3} presents the numerical calculation of localization properties in the presence of both random magnetic field and random potential. We show the dependence of $D_{1,2}$ and $KL$ on $\theta$ at a fixed total disorder $\rho_D=\sqrt{W^2+B^2}=2$. First note that at $\theta=\pi/2$, i.e. in the potential limit, $KL$ increases with the system size $L$ and both $D_1$ and $D_2$ decrease as $L$ grows. This behavior is an indication of the LO phase, which we expect.
This localization is well pronounced being compared to the extended phase, exhibited in purely magnetic disorder. This can be seen from Fig.~\ref{fig:2} and is also clear from the dependence of $D_{1,2}$ and $KL$ on $\theta$ close to $\theta=0$ in Fig.~\ref{fig:3}. It is
difficult to find any deviation of $KL$ from the ergodic value $KL=2$, whereas both $D_1$ and $D_2$ are close to $1$ and increase with $L$.

The most interesting and somewhat unexpected is the fact that this extended behavior persists at
$\rho_D=2$ for almost all values of $\theta$. Deviations are visible only when $\pi/2-\theta$ becomes smaller than $0.045 \pi$ (which corresponds to $B<0.3$ ). The behavior of $D_{1,2} $ and KL indicates the transition to the LO phase at $\theta^C \approx 0.497 \pi$ , i.e. $B^C \simeq 0.01$. This tiny field is sufficient
to destroy the LO phase created by the potential disorder $W \approx 2$. This can be seen from Fig.~\ref{fig:4} presented in terms of the potential disorder strength $W$ and magnetic field strength $B$. In particular, the LO phase can be recovered
by increasing the amplitude of the random magnetic field up to $B \simeq 8$. In the interval
$B^C < B \lesssim 8$ the eigenstates of the disordered Hamiltonian $H_0 + H_D$ are extended. It is possible that close to the edges of this interval the states loose ergodicity and become NEE. 
Unfortunately, the available sizes $L$ of the system for our computation are too small to justify
this guess. However, the reentrance to the localized regime with increasing $B$ is hard to dispute.

 For $\rho_D  \gg 1$ the extended phase is not expected, since both disorder components damp effective interparticle interactions~\cite{SupplMat}. From Fig.~\ref{fig:2} one expects this limit to be achived at $\rho_D \simeq 8.57 $.  In the other limit, $\rho_D  \ll 1$, the states are localized at $\theta = \frac{\pi}{2}$. Sufficiently weak random magnetic field induces effective interholon interaction, which completely delocalizes the states. This implies $\theta_c  \rightarrow  0.5 \pi $ with $\rho_D \rightarrow 0 $. Our final results for the phase diagram are presented in Fig.~\ref{fig:4}.


{\it Conclusions.} Compared to the potential static disorder, the random magnetic field has a dramatically different effect on the quantum many-body states of the 1D Hubbard model with infinitely strong on-site repulsion. The potential disorder does not violate the integrability of the problem - the holons remain free spinless fermions albeit subject to the random potential. Accordingly, the one-particle wave functions are localized at arbitrarily weak disorder. To complete the mapping onto the standard free fermion problem one has to take into account the quasiperiodic boundary conditions (QBC) for these functions. The QBC reflect the cyclical permutation in the spin sequence caused by the motion of the holons. Each stationary state of the system is characterized by a particular occupation of single-holon states as well as by the phase  $\phi$ of the QBC, which can take several values determined by the spin pattern.

 The infinite on-site repulsion preserves the spin pattern up to cyclical permutations irrespective of the type of disorder. In the potential case distinct cyclical transmutations do not change the potential energy giving rise the “quasimomentum”-phase $\phi$ of the QBC. Violation of this symmetry by the random magnetic field hybridizes the states with different $\phi$ and different holon quantum numbers making the system non-integrable and possibly extended in the MBL sense.
 
 A very strong disorder of any type practically excludes the cyclical transmutations and many-body states are localized in the computational basis. For a purely potential disorder it means that the eigenenergies are degenerate in $\phi$. If the potential disorder is weak but finite, a gradual reduction of the random magnetic field from a large value eventually causes the many-body delocalization. However, a further decrease should recover the localized phase.  
 
 These predictions are supported by the Bethe Ansatz based analytics~\cite{DenisWork} and clearly demonstrate the reentrant MBL behavior caused by the random magnetic field. Unfortunately, the size of the system available for the exact diagonalization turns out to be too small for the analysis of the ergodicity of the extended states.
 The Hamiltonian~(\ref{H_0}),(\ref{H_D}) can be implemented in systems of ultracold atoms or existing multi-qubit arrays. Being qualitatively transparent with well-developed analytical approaches, this model has advantages over e.g. spin chains as a laboratory for investigating MBL and searching for Quantum Supremacy~\citep{Arute2019}. 
 
 \begin{acknowledgments} 
We thank V.E. Kravtsov, V. Gritsev, O. Gamayun, and V. Smelyanskiy for useful discussions. This research was supported in part through computational resources of HPC facilities at NRU HSE and by the Russian Science Foundation Grant No. 20-42-05002. We also acknowledge support of this work by Rosatom.
 \end{acknowledgments}
   \bibliographystyle{apsrev4-1} 
\newpage
$~$
\newpage
\appendix
\setcounter{figure}{0}    
\setcounter{equation}{0}    

\renewcommand{\thefigure}{S\arabic{figure}}
\renewcommand{\theequation}{S\arabic{equation}}

\begin{widetext}
\begin{center}
\vskip 2cm
{\bf{\Large{Supplemental material }}}
\end{center}
\subsection{Exact diagonalization (ED) results of $D_1$ for various disorder amplitudes $\rho_D$}

To obtain a qualitative sketch of the phase diagram, we repeated our ED calculations of  $D_1$ for the system sizes $L=\left\lbrace 12,16 \right\rbrace$ and the values of $\rho_D=\left\lbrace 2,3,4,5,6,7,8,9,10,11 \right\rbrace$. Based on the following data, we sketched the phase diagram (in the thermodynamic limit), which is represented as~Fig.~\ref{fig:4} in the main text.
\begin{figure}[!htbp]
\includegraphics[width=18 cm]{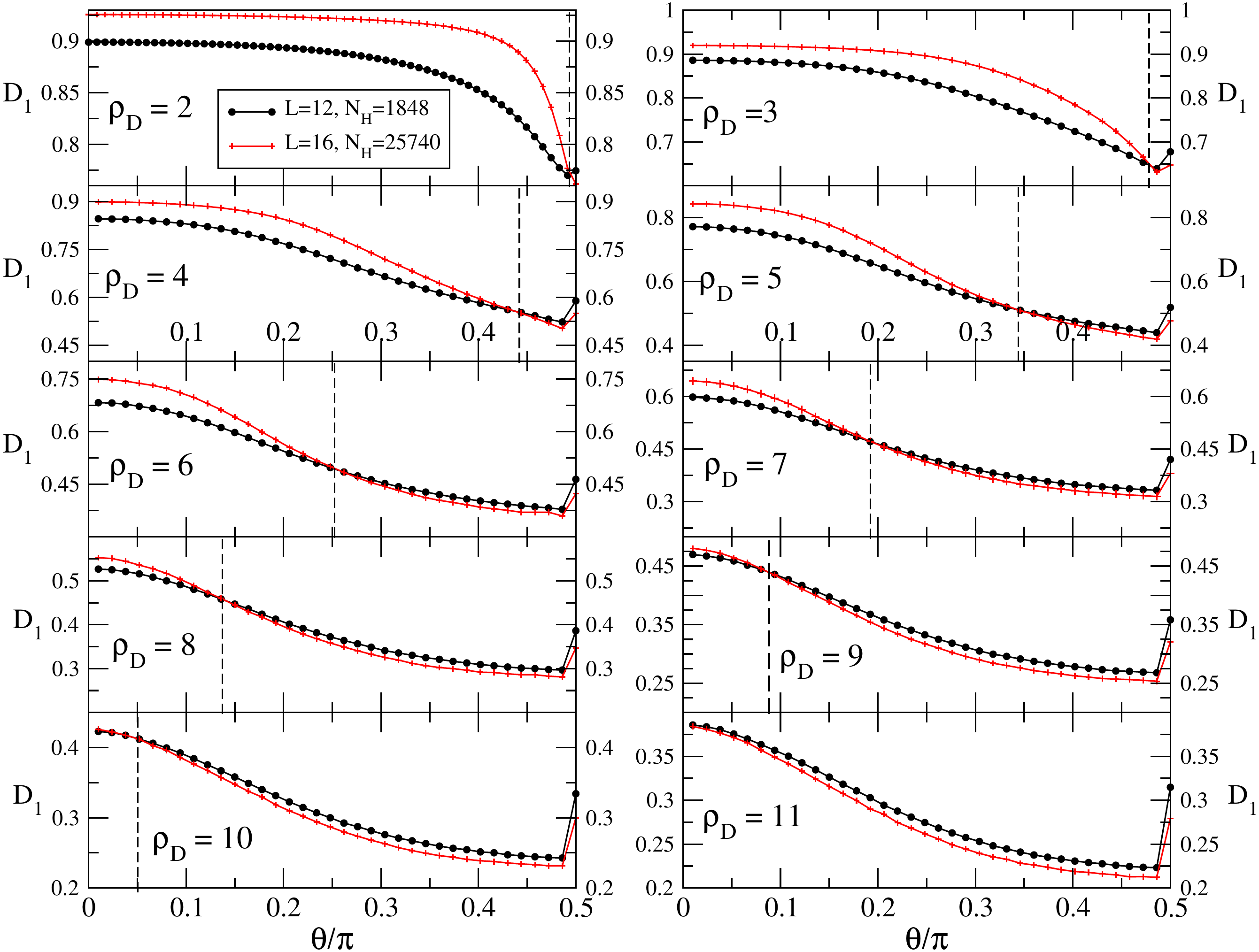}
\caption{ The fractal dimensions $D_1$ versus the angle~$\theta$ [see Eq.~(5) of the main text] for various values of $\rho_D$. The considered system sizes are $L = 12$ (black filled circles) and $L = 16 $ (red crosses). Error bars smaller than the symbol size are omitted. } 
\label{figS:1}
\end{figure} 

\end{widetext}
\end{document}